# AI-Based Reconstruction from Inherited Personal Data: Analysis, Feasibility, and Prospects

**Mark Zilberman, M.Sc.**
**Shiny World Corporation, Canada**
email: mzilberman137@gmail.com

## Abstract

This article explores the feasibility of creating an "electronic copy" of a deceased researcher by training artificial intelligence (AI) on the data stored in their personal computers. By analyzing typical data volumes on inherited researcher computers—including textual files such as articles, emails, and drafts—it is estimated that approximately one million words are available for AI training. This volume is sufficient for fine-tuning advanced pre-trained models like GPT-4 to replicate a researcher's writing style, domain expertise, and rhetorical voice with high fidelity. The study also discusses the potential enhancements from including non-textual data and file metadata to enrich the AI's representation of the researcher. Extensions of the concept include communication between living researchers and their electronic copies, collaboration among individual electronic copies, as well as the creation and interconnection of organizational electronic copies to optimize information access and strategic decision-making. Ethical considerations such as ownership and security of these electronic copies are highlighted as critical for responsible implementation. The findings suggest promising opportunities for AI-driven preservation and augmentation of intellectual legacy.

## Introduction

It is not uncommon for scientists, researchers, engineers, or other intellectuals to leave behind computers containing heterogeneous and unsystematized information after their death. These digital archives may include completed and draft versions of articles, letters (sent, received, or unfinished), screenshots, tables, bills, photographs, audio and video files, and other forms of data. A similar situation may arise when an individual permanently transitions to a different profession.

In most cases, such computers are either preserved as archives for potential future reference or eventually reformatted. However, the use of artificial intelligence (AI) has the potential to significantly transform this situation. It may become possible for the inheritor of a deceased intellectual's computer to create an 'electronic copy' of that individual by training AI models on the inherited digital content.

Throughout this article, we will refer to the original user of the computer as the "researcher," acknowledging that this individual may, in reality, be a scientist, engineer, or another form of intellectual.

**Data Lakes**

From a data architecture perspective, heterogeneous and unsystematized information stored on legacy computer systems can be described as a "data lake" (Sawadogo & Darmont, 2021). Table 1 below presents the key properties of data lakes. These properties illustrate why data lakes are often leveraged in modern analytics and machine learning workflows.

**Table 1**
*Key Properties of Data Lakes*

| Feature | Description |
|---|---|
| Raw data storage | Stores structured data (e.g., tables), semi-structured data (e.g., JSON, XML), and unstructured data (e.g., PDFs, videos, logs). |
| Schema-on-read | The structure is defined only at the time of data access or analysis, not during data storage. |
| Scalability | Capable of storing petabytes of data, often leveraging cloud platforms such as AWS S3, Azure Data Lake, or Hadoop-based systems. |
| Flexibility | Supports a wide range of use cases, including analytics, machine learning, and data science, without requiring predefined data models. |

**Note.** A key distinction between data located on inherited personal computers and modern data lakes lies in its *mono-source* origin. All content located on an individual's computer belongs to that specific user. While the user may not be the original author of every file (e.g., received emails or PDF documents authored by others), each file typically bears some relevance or connection to the user.

**Data Lakes and AI**

Data lakes are a common and strategically important resource for training artificial intelligence (AI) models (see Table 2).

**Table 2**
*How a Data Lake Supports AI Training*

| Role | How It Helps |
|---|---|
| Central storage | Collects and stores raw data from various sources (e.g., text, images, logs, tables, databases, audio, video), all of which can be valuable for AI training. |
| Scalability | Manages large volumes of data, which are often required to train effective AI models. |
| Flexible formats | Supports structured, semi-structured, and unstructured data, making it well-suited for the diverse input types used in AI applications. |

| Role | How It Helps |
|---|---|
| On-demand access | Enables retrieval of specific datasets or formats as needed, without the need to reprocess the entire data lake. |
| Versioning and governance | Tools such as AWS Glue, Microsoft Purview (formerly Azure Purview), and Delta Lake provide metadata management, data lineage, and access control, supporting data quality, traceability, and reproducibility. |

**Typical AI Workflow with a Data Lake**

1. Ingest raw data: Load unprocessed data—such as text, images, logs, tables, and audio—into the data lake.

2. Catalog and tag data: Use metadata management tools like Apache Hive Metastore or AWS Glue Data Catalog to organize and classify the ingested data.

3. Filter and preprocess datasets: Prepare the data for modeling by performing tasks such as text normalization, image resizing, or log parsing.

4. Export or stream processed data: Send preprocessed datasets to model training environments using platforms like PyTorch, TensorFlow, Amazon SageMaker, or Databricks.

5. Train and evaluate models: Use machine learning frameworks to train models on the curated datasets and assess their performance.

## Estimating the Data Volume Required for AI Training and the Data Available on Inherited Personal Computers

Before discussing the possibility of creating an electronic copy of a deceased researcher, it is important to assess the feasibility of such an endeavor. Do inherited computers of an average researcher typically contain enough data volume to effectively train an AI model? Additionally, what level of quality or fidelity can we realistically expect from an "electronic copy" produced through such training?

**Volume of Data Required for AI Training**

The volume of data needed for AI training depends on the chosen training method. There are two main approaches:

*a. Training from Scratch*

- Requires millions to billions of words.

- Is extremely resource-intensive and typically used by organizations such as OpenAI to develop general-purpose models.

- Is not practical for modeling an individual's profile.

*b. Fine-Tuning or Prompt Engineering*

- Is more feasible for creating personalized AI models.
- Involves using a pre-trained model (e.g., GPT-4) and fine-tuning it on a smaller, focused dataset (OpenAI, 2023a).

Table 3 below summarizes the approximate amount and quality of writing samples required for AI training. The data were provided by ChatGPT (OpenAI, 2023b).

**Table 3**
*Recommended Amount of Writing Samples for AI Training by Objective*

| Objective | Recommended sample size |
|---|---|
| Simulate style via prompting | Approximately 5–20 well-written samples (each can be a few paragraphs long) |
| Fine-tune a model | Approximately 10,000–50,000+ words (about 20–100+ pages) |
| Train from scratch | 100 million+ words (not recommended for style cloning) |

**Volume of Data on Inherited Personal Computers**

The amount of data available on an inherited personal computer can be estimated based on the research by Dinneen and Nguyen (2021). Their study analyzed 49 million files across 348 user collections, yielding an average of 140,804 files per user collection. The authors took additional precautions "to exclude files not managed by the participant: hidden files and common folders containing operating system files were explicitly ignored" (Dinneen & Nguyen, 2021).

The percentage of textual files is reported in the research by Shelly (2024), who analyzed the distribution of files in the Special Collections Research Center at North Carolina State University, USA, which has collected and preserved more than 10 terabytes of born-digital materials (see Table 4).

**Table 4**
*Types and Percentage of Textual Files in the Special Collections Research Center at North Carolina State University, USA*

| Document format | Percentage |
|---|---|
| PDF 1.4 | 2% |
| PDF 1.3 | 1% |
| Plain text | 3% |
| MS Word | 4% |
| MS Word for Windows | 1% |
| **Total textual documents** | **11 %** |

By combining the findings of Shelly (2024) and Dinneen and Nguyen (2021), we estimate the number of textual files present on an inherited researcher's computer (see Table 5).

**Table 5**
*Estimated Number of Textual Files on a Researcher's Computer*

| Document format | Percentage | Number of files |
|---|---|---|
| PDF 1.4 | 2% | 2,800 |
| PDF 1.3 | 1% | 1,400 |
| Plain text | 3% | 4,200 |
| MS Word | 4% | 5,600 |
| MS Word for Windows | 1% | 1,400 |
| **Total textual documents** | **11%** | **15,400** |

Choueiry (2022) analyzed a random sample of 61,519 full-text research papers uploaded to PubMed Central between 2016 and 2021 and reported that the mean length of a research paper is 4,539 words.

Therefore, the total number of words in textual files on a researcher's computer can be roughly estimated as 15,400 × 4,539 = 69,900,600 words. Obviously, the researcher is not the sole author of all documents stored on their computer; many documents may have been written by colleagues, students, or downloaded from the Internet. Since not all of these documents are scientific articles, their lengths may vary and are not necessarily around 4,500 words. However, the presence of these documents on the researcher's computer still reflects their interests, which can be valuable for AI training aimed at creating an "electronic copy" closely resembling the original.

**Estimation of the Number of Words in Researcher's Articles**

Ioannidis, Baas, Klavans, and Boyack (2019) showed that the number of lifetime publications for most productive scientists ranges between 100 and 300, with an average of 209 publications. Considering the average length of a research paper as 4,539 words (Choueiry, 2022), we can roughly estimate the total number of words in articles published by a researcher as 209 × 4,539 = 948,651 words. Adding the words from sent and drafted but unsent emails, we arrive at approximately 1,000,000 words attributed to the researcher.

According to ChatGPT (OpenAI, 2023b), "with a 1,000,000-word dataset (~2,000–4,000 pages), you're in a very strong position to train an AI that closely mimics someone's writing style, tone, and even reasoning patterns — especially if the data is consistent and well-organized."

With a training database of about 1,000,000 words, AI can:

- Ensure high-quality style mimicry:
  The AI can convincingly replicate vocabulary, sentence structure, tone, rhythm, and typical expressions. If the writing includes dialogue, it may even emulate the person's manner of speaking.

- Ensure topic familiarity:
  If the text covers specific domains (e.g., science, culture, education), the AI can "learn" to respond within those domains with high confidence and authenticity.

- Support personality and voice:
  Given enough examples of opinions, argument structures, and rhetorical habits, the AI can often approximate the person's unique voice in new responses.

  However, there are certain limitations for AI trained on a dataset of about 1,000,000 words:

- Reasoning does not imply consciousness:
  The AI will mimic style and typical thought patterns but does not possess the person's real judgment, intent, or beliefs. While it may simulate how the person might respond to new topics, it cannot truly "think like them" beyond the patterns it has observed.

- Context gaps:
  If topics outside the scope of the 1 million-word dataset arise, the AI may maintain the style but lack sufficient content knowledge.

- Consistency requires structure:
  The quality of the resulting "electronic copy" depends not only on the word count but also on how well the data is curated—for example, whether it is organized by topic, tone, or format.

Table 6 presents a summary of training, fine-tuning, and prompting methods based on a 1,000,000-word dataset. The estimations were generated by ChatGPT AI.

**Table 6**

*Summary of Training, Fine-Tuning, and Prompting Methods Based on a 1,000,000-Word Dataset*

| Method | Feasibility with 1M words | Output quality |
|---|---|---|
| Prompt-based emulation | Good for style-focused tasks | Good to excellent |
| Fine-tuning a GPT model | Excellent — 1M words is ideal | High (style, topic, and tone mimicry) |
| Training a model from scratch | Still too little | Low unless using a small language model base |

As observed, training an AI model from scratch is not feasible with a dataset of one million words; however, fine-tuning a GPT model can yield high-quality results.

Given that we can rely on approximately 1,000,000 words authored by the researcher (sometimes with co-authors), along with about 70,000,000 words from textual files stored on the researcher's computer, the creation of an electronic copy of a deceased researcher using AI trained on data from their personal computer(s) *appears feasible*.

**Non-Textual Data**

So far, we have estimated the volume of only the textual training dataset. However, according to Dinneen and Nguyen (2021), 89% of files on a researcher's PC are non-textual. These include images, photos, videos, audio recordings, and more. The generation of an "electronic copy" of a deceased researcher by AI would be significantly enhanced by including and analyzing non-textual files during training and testing. Incorporating such files could improve the richness of the researcher's biography, provide deeper insights into their thoughts, and capture the emergence and decline of their interests.

**Metadata**

File creation dates—and other metadata—should also be taken into account to help the AI understand the progression of the deceased researcher's ideas and biographical timeline.

According to ChatGPT,

> "AI models like mine are trained primarily on the *textual content* of data, not on file metadata such as creation dates, author names, or timestamps — unless that information is explicitly included in the text itself. For example, if an article says "As of April 2020, we know…" then that date becomes part of the training data. But if a file's metadata says it was created in April 2020 and the text doesn't reference that, the model won't be aware of it."

When training AI to create an "electronic copy" of a deceased researcher, it is important to incorporate available metadata, such as file creation and modification dates, authorship information, photo metadata (e.g., camera type, geotags), metadata from PDF files, etc.

## Non-Technical Considerations and Perspectives

**Ownership of the Created "Electronic Copy" of a Deceased Researcher**

While the creation of an "electronic copy" of a deceased researcher appears technically feasible, the question of ownership of the resulting digital entity is less clear. As an initial assumption, ownership would likely belong to the individual who inherited the researcher's computer(s). However, complexities arise if the researcher used multiple computers that were inherited by different people or organizations—for example, a home computer versus a work computer. This legal and ethical issue requires careful consideration and may warrant separate, detailed analysis.

**Communication of Researcher with Their Electronic Copy**

An interesting opportunity arises when AI is trained on the data of a *living* researcher, enabling the researcher to communicate with their own "electronic copy." While this concept may seem unusual, such interaction has the potential to significantly enhance the researcher's productivity.

- Instead of recalling details or performing time-consuming searches through files, the researcher can simply request specific information or file locations from their electronic copy.

- The researcher may forget or overlook certain ideas or facts that could support or guide future research; the electronic copy can remind them of these ideas and facts from their own or others' publications stored on their computer.

- The electronic copy can analyze large volumes of information on the researcher's machine and reveal correlations, trends, and connections that the researcher may not have previously considered.

**Collaboration of "Electronic Copies" of Individual Researchers**

So far, we have discussed AI training on computer(s) belonging to a single researcher. However, under certain circumstances, individual researchers may collaborate by combining their AI training datasets for mutual benefit. Multiple researchers could:

- Provide data from their personal computers to train a single, unified AI, creating an electronic copy that represents the collective knowledge and style of the research group.

- Train electronic copies individually on their own computers but enable communication between these copies, establishing security protocols and frameworks for information exchange among the individual electronic copies.

**"Electronic Copy of Organization"**

Training AI on dedicated—or all—computers within an entire organization may create what can be termed an "electronic copy of the organization." Such an electronic copy could significantly enhance and accelerate access to documents and information used by departments and individual members. Furthermore, this organizational electronic copy may suggest solutions and optimizations that individual members might not have considered, due to limited time or insufficient access to relevant information. Naturally, security protocols governing access to and use of such an electronic copy must be carefully designed and rigorously implemented in advance.

**Collaboration Between Electronic Copies of Organizations**

Similar to the collaboration between electronic copies of individual researchers, the electronic copies of organizations may communicate and collaborate to implement common strategies and mutually benefit from shared insights. Such collaboration could also reveal solutions and optimizations that individual members of the organizations might not have considered, due to limited time or access to necessary information. As with previous examples, security measures governing access to and use of these collaborative electronic copies must be carefully designed and implemented in advance.

**Conclusion**

Productive researchers generate approximately 1,000,000 words across their original articles, memos, and emails throughout their careers. This volume of text is sufficient to train an AI model capable of becoming an "electronic copy" of a deceased researcher, with the ability to:

- replicate vocabulary, sentence structure, tone, rhythm, and typical expressions convincingly;

- respond confidently and authentically within the researcher's professional domain;

- approximate the individual's voice in new responses, given enough examples of opinions, argument structures, and rhetorical habits.

Extensions of this approach may include communication between a living researcher and their own electronic copy, collaboration among electronic copies of individual researchers, the creation of electronic copies of organizations, and collaboration among these organizational copies for mutual benefit, among many other potential applications.

**References**

Comeau, D. C., Wei, C.-H., Islamaj Doğan, R., & Lu, Z. (2019). PMC text mining subset in BioC: About three million full-text articles and growing. *Bioinformatics, 35*(18), 3533–3535.
https://doi.org/10.1093/bioinformatics/btz070

Dinneen, J. D., Nguyen, B. X. (2021). How Big Are Peoples' Computer Files? File Size Distributions Among User Managed Collections. In *ASIS&T '21: Proceedings of the 84th Annual Meeting of the Association for Information Science & Technology, 58* https://doi.org/10.1002/pra2.472

Choueiry, G. (2022). How long should a research paper be? Data from 61,519 examples. *Quantifying Health*. https://quantifyinghealth.com/length-of-a-research-paper

Ioannidis, J. P. A., Baas, J., Klavans, R., & Boyack, K. W. (2019). A standardized citation metrics author database annotated for scientific field. *PLOS Biology, 17*(8), e3000384. https://doi.org/10.1371/journal.pbio.3000384

OpenAI. (2023a). GPT-4 technical report. https://openai.com/research/gpt-4

OpenAI. (2023b). ChatGPT [Large language model]. https://chat.openai.com/

Sawadogo, P., Darmont, J. On data lake architectures and metadata management. *J Intell Inf Syst* **56**, 97–120 (2021). https://doi.org/10.1007/s10844-020-00608-7

Shelly, B. (2024). *File formats count: Results of a survey*. North Carolina State University Libraries. https://www.lib.ncsu.edu/news/special-collections/file-formats-count-results-survey